\documentclass{article}[10pt]
\usepackage{amsfonts}
\usepackage{amssymb}
\usepackage{graphicx}
\usepackage{psfrag}
\usepackage{natbib}

\begin{document}
\title{Material forces in the context of biotissue remodelling}
\author{K. Garikipati\thanks{Asst. Professor, Department of Mechanical Engineering, {\tt
krishna@umich.edu}}, H. Narayanan\thanks{Graduate research
assistant, Department of Mechanical Engineering}, E. M.
Arruda\thanks{Assoc. Professor, Department of Mechanical
Engineering and Program in Macromolecular Science and
Engineering}, K. Grosh\thanks{Assoc. Professor, Departments of
Mechanical Engineering, and Biomedical Engineering}, S.
Calve\thanks{Graduate research assistant, Program in
Macromolecular Science and
Engineering}\\
University of Michigan, Ann Arbor, Michigan 48109, USA}
\date{}
\maketitle
%
\font\title=cmbx10 scaled \magstep2 \font\bigit=cmti10
\font\mbold=cmssbx10 \font\af=cmcsc10 \font\aff=cmr8
\font\contents=cmr10 at 10 truept \font\bc=cmbx10 at 11 truept
\font\ninerm=cmr9 \font\bbk=cmsy10 \font\ninebi=cmmib10 at 7pt
\font\bb=cmssbx10
\def\slsmall{\small}
\def\slb{\textbf}
\def\sltt{\tt}
\def\slcal{\cal}
\def\slsl{\textsl}
\def\slnames{\textsc}
%
%
\font\lesshuge = cmbx10 scaled \magstep 2 \font\big  = cmbx10
scaled \magstep 1 \font\med  = cmr10
\font\smallb = cmbx10
\font\names = cmcsc10 \font\addrs = cmti10
%
%
\def\Jmat{{\hbox{\bb J}}}
\def\unitcirc{{\hbox{\cal C}}}
\def\tF{{\hbox{\tiny F}}}
\def\tI{{\hbox{\tiny I}}}
\def\dt{{\Delta t}}
\def\n1{{n+1}}
\def\nh{{n+{1\over 2}}}
\def\nnode{{n_{\rm node}}}
\def\Gd{{\Gamma_{\varphi}}}
\def\Gt{{\Gamma_{\sigma}}}
\def\ll{\big\langle}
\def\rr{\big\rangle}
\def\Ll{\Big\langle}
\def\Rr{\Big\rangle}
\def\Pint{{{\cal P}_t^{\rm int}}}
\def\Dint{{{\cal D}_t^{\rm int}}}
\def\Vint{{{\it  V}_t^{\rm int}}}
\def\Vext{{{\it  V}_t^{\rm ext}}}
\def\Etot{{{\it  H}_t}}
\def\Pext{{{\cal P}_t^{\rm ext}}}
\def\Fext{{{\bF}_t^{\rm ext}}}
\def\Text{{{\bT}_t^{\rm ext}}}
\def\FextA{{{\bF}_\nh^{{\rm ext} A}}}
\def\FintA{{{\bF}_\nh^{{\rm int} A}}}
\def\e1{P1$\oplus$Bub\/P1\/E1}
\def\e2{P1$\oplus$Bub\/P1\/E2}
\def\h{\Delta t}
\def\a{\alpha}
\def\y{{\by}}
\def\x{{\bx}}
\def\z{{\bz}}
\def\ya{{\by}_{n+\alpha}}
\def\ytn{{\by}(t_n)}
\def\yn{{\by}_{n}}
\def\ye{{\by}_{n+1}}
\def\ta{t_{n+\alpha}}
\def\tn{t_{n}}
\def\te{t_{n+1}}
\def\f{{\bff}}
\def\fn{{\bff}\big({\by}(t_n)\big)}
\def\fa{{\bff}\big({\by}(t_{n+\alpha})\big)}
\def\yaa{{\by}_{n+(1-\alpha)}}
\def\taa{t_{n+(1-\alpha)}}

\def\Dt{{\Delta t}}
\def\Etoto{{{\it  H}_0}}
\def\Etotn{{{\it  H}_n}}
\def\Etotnn{{{\it H}_{n+1}}}
\def\Vintn{{{\it V}^{\rm int}_n}}
\def\Vintnn{{{\it V}^{\rm int}_{n+1}}}
\def\Vextn{{{\it V}^{\rm ext}_n}}
\def\Vextnn{{{\it V}^{\rm ext}_{n+1}}}
\def\DDint{{\Delta{\cal D}^{\rm int}_{[t_n,t_\n1]}}}

\def\tr{\mathop{\rm tr}\nolimits}
\def\cay{\mathop{\rm cay}\nolimits}
\def\exp{\mathop{\rm exp}\nolimits}
\def\trn{^{\hbox{\tiny T}}}
\def\and{\quad{\rm and}\quad}
\def\inn{\quad{\rm in}\quad}
\def\where{\quad{\rm where}\quad}
\def\htheta{{\theta \over 2}}
\def\bigbullet{\raise 0.7pt \hbox{$\bigoldbullet$}}
%
%
\def\br{\Bvarphi}
\def\and{\quad{\rm and}\quad}
\def\ce{\centerline}
\def\ul{\underbar}
\def\bg{{\bi g}}
\def\norm{|\! |\! |}

\def\II{\hbox{\bbk I}}
\def\IJ{\hbox{\bbk J}}
\def\IL{\hbox{\bbk L}}
\def\IQ{\hbox{\bbk Q}}

\def\ic {\hbox{\bb c}}
\def\ID {\hbox{\bb D}}
\def\IH {\hbox{\bb H}}
\def\IR {\hbox{\bb R}}

\def\IC {\hbox{\gm C}}
\def\IN {\hbox{\gm N}}
\def\IB {\hbox{\gm B}}
\def\IG {\hbox{\gm G}}
\def\IF {\hbox{\gm F}}
\def\IK {\hbox{\gm K}}
\def\IE {\hbox{\gm E}}
\def\IS {\hbox{\gm S}}
\def\Ih {\hbox{\gm h}}

\def\In {{\hbox{{\bb n}}}}
\def\Ir {{\hbox{{\bb r}}}}
\def\Ib {{\hbox{{\bb b}}}}
\def\Im {{\hbox{{\bb m}}}}
\def\bone{{\bf 1}}
\def\bzero{{\bf 0}}

\def\bcf  {{\hbox{{\bb c}}}_f}
\def\bcg  {{\hbox{{\bb c}}}_g}
\def\bcfnn{{\hbox{{\bb c}}}_{f\,n+1}}
\def\bcgnn{{\hbox{{\bb c}}}_{g\,n+1}}
\def\bcfn {{\hbox{{\bb c}}}_{f\,n}}
\def\bcgn {{\hbox{{\bb c}}}_{g\,n}}
\def\bc {\bi c}

\def\Ief   {{\hbox{{\bb e}}}_f}
\def\Ieg   {{\hbox{{\bb e}}}_g}
\def\Iefnn {{\hbox{{\bb e}}}_{f\,n+1}}
\def\Iegnn {{\hbox{{\bb e}}}_{g\,n+1}}
\def\Iefnb {{\hbox{{\bb e}}}_{f\,n+\beta}}
\def\Iegnb {{\hbox{{\bb e}}}_{g\,n+\beta}}
\def\Iefnnb{{\hbox{{\bb e}}}_{f\,n+(1-\beta)}}
\def\Iegnnb{{\hbox{{\bb e}}}_{g\,n+(1-\beta)}}
\def\Iefna {{\hbox{{\bb e}}}_{f\,n+\vartheta}}
\def\Iegna {{\hbox{{\bb e}}}_{g\,n+\vartheta}}
\def\Iefn  {{\hbox{{\bb e}}}_{f\,n}}
\def\Iegn  {{\hbox{{\bb e}}}_{g\,n}}

\def\If {\hbox{{\bb f}}}
\def\Ig {\hbox{{\bb g}}}

\def\half{{\textstyle{1 \over 2}}}
\def\third{{\textstyle{1 \over 3}}}
\def\fourth{{\textstyle{{1 \over 4}}}}
\def\twothird{{\textstyle {{2 \over 3}}}}
\def\ndim{{n_{\rm dim}}}
\def\nint{n_{\rm int}}
\def\lint{l_{\rm int}}
\def\nel{n_{\rm el}}
\def\DIV {\hbox{\rm div}}
\def\Grad{\hbox{\rm Grad}}
\def\sym{\mathop{\rm sym}\nolimits}
\def\dev{\mathop{\rm dev}\nolimits}
\def\Dev{\mathop {\rm DEV}\nolimits}
\def\bfb {{\bi b}}
\def\Bnabla{\nabla}
\def\bG{{\bi G}}
\def\jmpdelu{{\lbrack\!\lbrack \Delta u\rbrack\!\rbrack}}
\def\jmpudot{{\lbrack\!\lbrack\dot u\rbrack\!\rbrack}}
\def\jmpu{{\lbrack\!\lbrack u\rbrack\!\rbrack}}
\def\jmphi{{\lbrack\!\lbrack\varphi\rbrack\!\rbrack}}
\def\ljmp{{\lbrack\!\lbrack}}
\def\rjmp{{\rbrack\!\rbrack}}
\def\sign{{\rm sign}}
\def\nn{{n+1}}
\def\na{{n+\vartheta}}
\def\nna{{n+(1-\vartheta)}}
\def\nt{{n+{1\over 2}}}
\def\nb{{n+\beta}}
\def\nbb{{n+(1-\beta)}}
\def\sB{{\cal B}}
\def\sU{{\cal U}}
\def\sC{{\cal C}}
\def\sS{{\cal S}}
\def\sV{{\cal V}}
\def\sL{{\cal L}}
\def\sO{{\cal O}}
\def\sH{{\cal H}}
\def\sG{{\cal G}}
\def\sM{{\cal M}}
\def\sN{{\cal N}}
\def\sF{{\cal F}}
\def\sW{{\cal W}}
\def\sT{{\cal T}}
\def\sR{{\cal R}}
\def\sJ{{\cal J}}
\def\sK{{\cal K}}

\def\overstuff#1#2{\vbox{\ialign{##\crcr
$\hfil\scriptstyle#1\hfil$\crcr\noalign{\kern1pt\nointerlineskip}
$\hfil\displaystyle{#2}\hfil$\crcr}}}
\def\starr   {\delta\br}
\def\stard   {\delta\bd}
\def\starD   {\delta\bD}
\def\starp   {\delta\bp}
\def\starpi  {\delta\Bpi}
\def\starPhi {\delta\Phi}
\def\starPi  {\delta\Pi}
\def\starP   {\delta P}
\def\starZ   {\delta Z}
\def\starphi {\overstuff {*}{\br}}
\def\mt#1{{\scriptstyle #1}}
\mathchardef \BG = "0947 \mathchardef \BC = "080D
\mathchardef\bigoldbullet= "220F
\def\MK{\bM\!\bK}
\def\GK{\BG\!\bK}
\def\sljb{ {{\sl j}\,}\! }
\def\slJb{ {{\sl J}\,}\! }
\def\slj {{\sl j}}

\def\starrh   {\delta\br^h}
\def\stardh   {\delta\bd^h}
\def\starPhih  {\delta\Phi^h}

\def\Gdyn {G_{\rm dyn}(Z;\starPhi)}
\def\Gstat{G_{\rm stat}(\Phi;\starPhi)}
\def\Dp{\bp_{n+1}-\bp_n}
\def\Dpi{\Bpi_{n+1}-\Bpi_n}
\def\Gstath{G_{\rm stat}(\Phi_n,\Phi_\nn;\starPhi)}
\def\Gdynh {G_{\rm dyn}(Z_n,Z_{n+1};\starPhi)}
\def\Gstathh{G_{\rm stat}(\Phi_n^h,\Phi_\nn^h;\starPhih)}
\def\Gdynhh {G_{\rm dyn}(Z_n^h,Z_{n+1}^h;\starPhi^h)}
\def\Dph{\bp_{n+1}^h-\bp_n^h}
\def\Dpih{\Bpi_{n+1}^h-\Bpi_n^h}
\font\gm=cmsy10
\def\Iinf     {\hbox{\gm I}_\infty}
\def\Reals    {\hbox{\gm R}}
\def\Complexes{\hbox{\gm C}}
\def\IC       {\hbox{\gm C}}
\def\IG       {\hbox{\gm G}}
\def\IM       {\hbox{\gm M}}
\def\IN       {\hbox{\gm N}}
\def\IX       {\hbox{\gm X}}
\def\IF       {\hbox{\gm F}}
\def\IE       {\hbox{\gm E}}
\def\IS       {\hbox{\gm S}}
\def\II       {\hbox{\gm I}}
\def\IC       {\hbox{\gm C}}
\def\IH       {\hbox{\gm H}}
\def\itIM     {\hbox{\gm M}}
\def\IP       {\hbox{\gm P}}
\def\I0       {\hbox{\gm O}}
\def\bzero    {\hbox{\bf 0}}
\def\bbE      {\hbox{\gm E}}
\def\bbF      {\hbox{\gm F}}
\def\bbG      {\hbox{\gm G}}
\def\bbM      {\hbox{\gm M}}
\def\bk       {{\bi k}}
\def\sE       {{\cal E}}
\def\sS       {{\cal S}}
\def\sH       {{\cal H}}
\def\sD       {{\cal D}}
\def\sG       {{\cal G}}
\def\sP       {{\cal P}}
\def\bp       {{\bi p}}
\def \ce    {\centerline    }
\def \ni    {\noindent      }
\def \lp    {\par\noindent  }
\def \ul    {\underbar      }
\def\BGamma{\mbox{\boldmath$\Gamma$}}
\def\BDelta{\mbox{\boldmath$\Delta$}}
\def\BTheta{\mbox{\boldmath$\Theta$}}
\def\BLambda{\mbox{\boldmath$\Lambda$}}
\def\BXi{\mbox{\boldmath$\Xi$}}
\def\BPi{\mbox{\boldmath$\Pi$}}
\def\BSigma{\mbox{\boldmath$\Sigma$}}
\def\BUpsilon{\mbox{\boldmath$\Upsilon$}}
\def\BPhi{\mbox{\boldmath$\Phi$}}
\def\BPsi{\mbox{\boldmath$\Psi$}}
\def\BOmega{\mbox{\boldmath$\Omega$}}
\def\Balpha{\mbox{\boldmath$\alpha$}}
\def\Bbeta{\mbox{\boldmath$\beta$}}
\def\Bgamma{\mbox{\boldmath$\gamma$}}
\def\Bdelta{\mbox{\boldmath$\delta$}}
\def\Bepsilon{\mbox{\boldmath$\epsilon$}}
\def\Bzeta{\mbox{\boldmath$\zeta$}}
\def\Beta{\mbox{\boldmath$\eta$}}
\def\Btheta{\mbox{\boldmath$\theta$}}
\def\Biota{\mbox{\boldmath$\iota$}}
\def\Bkappa{\mbox{\boldmath$\kappa$}}
\def\Blambda{\mbox{\boldmath$\lambda$}}
\def\Bmu{\mbox{\boldmath$\mu$}}
\def\Bnu{\mbox{\boldmath$\nu$}}
\def\Bxi{\mbox{\boldmath$\xi$}}
\def\Bpi{\mbox{\boldmath$\pi$}}
\def\Brho{\mbox{\boldmath$\rho$}}
\def\Bsigma{\mbox{\boldmath$\sigma$}}
\def\Btau{\mbox{\boldmath$\tau$}}
\def\Bupsilon{\mbox{\boldmath$\upsilon$}}
\def\Bphi{\mbox{\boldmath$\phi$}}
\def\Bchi{\mbox{\boldmath$\chi$}}
\def\Bpsi{\mbox{\boldmath$\psi$}}
\def\Bomega{\mbox{\boldmath$\omega$}}
\def\Bvarepsilon{\mbox{\boldmath$\varepsilon$}}
\def\Bvartheta{\mbox{\boldmath$\vartheta$}}
\def\Bvarpi{\mbox{\boldmath$\varpi$}}
\def\Bvarrho{\mbox{\boldmath$\varrho$}}
\def\Bvarsigma{\mbox{\boldmath$\varsigma$}}
\def\Bvarphi{\mbox{\boldmath$\varphi$}}
\def\bA{\mbox{\boldmath$ A$}}
\def\bB{\mbox{\boldmath$ B$}}
\def\bC{\mbox{\boldmath$ C$}}
\def\bD{\mbox{\boldmath$ D$}}
\def\bE{\mbox{\boldmath$ E$}}
\def\bF{\mbox{\boldmath$ F$}}
\def\bG{\mbox{\boldmath$ G$}}
\def\bH{\mbox{\boldmath$ H$}}
\def\bI{\mbox{\boldmath$ I$}}
\def\bJ{\mbox{\boldmath$ J$}}
\def\bK{\mbox{\boldmath$ K$}}
\def\bL{\mbox{\boldmath$ L$}}
\def\bM{\mbox{\boldmath$ M$}}
\def\bN{\mbox{\boldmath$ N$}}
\def\bO{\mbox{\boldmath$ O$}}
\def\bP{\mbox{\boldmath$ P$}}
\def\bQ{\mbox{\boldmath$ Q$}}
\def\bR{\mbox{\boldmath$ R$}}
\def\bS{\mbox{\boldmath$ S$}}
\def\bT{\mbox{\boldmath$ T$}}
\def\bU{\mbox{\boldmath$ U$}}
\def\bV{\mbox{\boldmath$ V$}}
\def\bW{\mbox{\boldmath$ W$}}
\def\bX{\mbox{\boldmath$ X$}}
\def\bY{\mbox{\boldmath$ Y$}}
\def\bZ{\mbox{\boldmath$ Z$}}
\def\ba{\mbox{\boldmath$ a$}}
\def\bb{\mbox{\boldmath$ b$}}
\def\bc{\mbox{\boldmath$ c$}}
\def\bd{\mbox{\boldmath$ d$}}
\def\be{\mbox{\boldmath$ e$}}
\def\bff{\mbox{\boldmath$ f$}}
\def\bg{\mbox{\boldmath$ g$}}
\def\bh{\mbox{\boldmath$ h$}}
\def\bi{\mbox{\boldmath$ i$}}
\def\bj{\mbox{\boldmath$ j$}}
\def\bk{\mbox{\boldmath$ k$}}
\def\bl{\mbox{\boldmath$ l$}}
\def\bm{\mbox{\boldmath$ m$}}
\def\bn{\mbox{\boldmath$ n$}}
\def\bo{\mbox{\boldmath$ o$}}
\def\bp{\mbox{\boldmath$ p$}}
\def\bq{\mbox{\boldmath$ q$}}
\def\br{\mbox{\boldmath$ r$}}
\def\bs{\mbox{\boldmath$ s$}}
\def\bt{\mbox{\boldmath$ t$}}
\def\bu{\mbox{\boldmath$ u$}}
\def\bv{\mbox{\boldmath$ v$}}
\def\bw{\mbox{\boldmath$ w$}}
\def\bx{\mbox{\boldmath$ x$}}
\def\by{\mbox{\boldmath$ y$}}
\def\bz{\mbox{\boldmath$ z$}}
\begin{abstract}
Remodelling of biological tissue, due to changes in
microstructure, is treated in the continuum mechanical setting.
Microstructural change is expressed as an evolution of the
reference configuration. This evolution is expressed as a
point-to-point map from the reference configuration to a
remodelled configuration. A ``preferred'' change in configuration
is considered in the form of a globally incompatible tangent map.
This field could be experimentally determined, or specified from
other insight. Issues of global compatibility and evolution
equations for the resulting configurations are addressed. It is
hypothesized that the tissue reaches local equilibrium with
respect to changes in microstructure. A governing differential
equation and boundary conditions are obtained for the
microstructural changes by posing the problem in a variational
setting. The Eshelby stress tensor, a separate configurational
stress, and thermodynamic driving (material) forces arise in this
formulation, which is recognized as describing a process of
self-assembly. An example is presented to illustrate the
theoretical framework.
\end{abstract}

\section{Introduction}\label{sect1}

The development of biological tissue consists of distinct
processes of growth, remodelling and morphogenesis---a
classification suggested by \cite{Taber:95}. In our treatment of
the problem, growth is defined as the addition or depletion of
mass through processes of transport and reaction coupled with
mechanics. As a result there is an evolution of the concentrations
of the various species that make up the tissue. Nominally, these
include the solid phase (cells and extra cellular matrix), the
fluid phase (interstitial fluid), various amino acids, enzymes,
nutrients, and byproducts of reactions between them. The stress
and deformation state of the tissue also evolve due to mechanical
loads, and the coupling between transport, reaction and mechanics.

Remodelling is the process of microstructural reconfiguration
within the tissue. It can be viewed as an evolution of the
reference configuration to a ``remodelled'' configuration. While
it usually occurs simultaneously with growth, it is an independent
process. For the purpose of conceptual clarity we will ignore
growth in this paper, and focus upon a continuum mechanical
treatment of remodelling.

\begin{figure}[ht]
\unitlength1cm
\begin{minipage}[t]{5cm}
{\includegraphics[width=5cm]{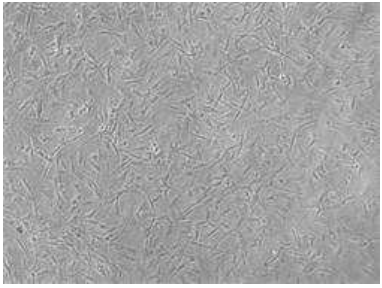}} \centerline{(a)}
\end{minipage}
\hfill
\begin{minipage}[t]{5.6cm}
{\includegraphics[width=5.6cm]{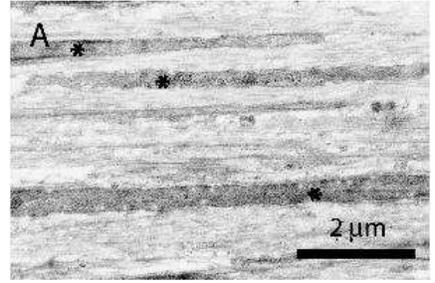}} \centerline{(b)}
\end{minipage}
\caption{(a) Micrograph taken 3 days after plating of cells, shows
a random distribution and orientation of tendon fibroblast cells
in engineered tendon. (b) As growth occurs the cells organize into
a more ordered microstructure seen in this micrograph taken about
a month after plating of cells. The horizontal alignment of cells
corresponds to the orientation of a uniaxial stress that was
imposed externally on the growing tendon construct. The alignment
of cells along the stress axis is evidence of remodelling due to
stress in the engineered tendon during growth.} \label{fig1}
\end{figure}

The microstructural reconfiguration that underlies remodelling is
a motion of material points in \emph{material space}. An example
of remodelling driven by stress is provided by the micrographs of
Figure \ref{fig1} from \cite{Calveetal:03}.

Remodelling also can be driven by the local density of the
tissue's solid or fluid phases, availability of various chemical
factors, temperature, etc. We assume that it is possible (through
experiments or other approaches) to define a phenomenological law
that specifies this evolution. Since any conditions that could
drive remodelling vary pointwise through the tissue, such a
``preferred'' remodelled state will, in general, be globally
incompatible. However, in its final remodelled state, the tissue
is virtually always free of such incompatibilities (see Figure
\ref{fig1}). We therefore propose that a further,
compatibilty-restoring material motion occurs, carrying material
points to the remodelled configuration.

\cite{TaberHumphrey:01} and \cite{AmbrosiMollica:02} have
previously referred to remodelling. However, the treatments in
these papers are based upon concentration (or density) changes in
growing tissue and the mechanics---mainly internal stress---that
is associated with them. By our definitions, these papers describe
growth rather than remodelling. In a largely descriptive paper,
\cite{HumphreyRajagopal:02} have proposed the evolution of
``natural configurations'' that seems closest to our ideas. To our
knowledge, however, no quantitative treatment exists paralleling
the ideas of microstructural reconfiguration, material motion,
material/configurational forces and their relation to remodelling,
as described in the present paper.

\section{Variational formulation: Material motion and material
forces}\label{sect2}

\begin{figure}[ht]
\psfrag{A}{\small $\Omega_0$} \psfrag{B}{\small $\Omega^\ast$}
\psfrag{C}{\small $\Omega_t$} \psfrag{D}{\small $\Bvarphi$}
\psfrag{G}{\small $\bu^\ast$} \psfrag{E}{\small $\Bkappa$}
\centering{\includegraphics[width=10cm]{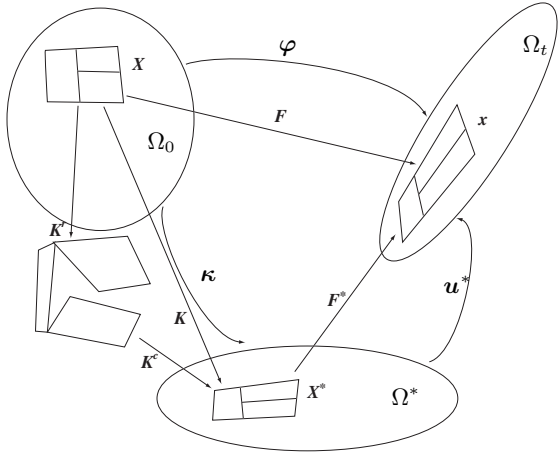}} \vskip -1cm
\caption{The kinematics of remodelling}\label{fig2}
\end{figure}

Figure \ref{fig2} depicts the kinematics associated with
remodelling. The preferred remodelled state is given by a tangent
map of material motion $\bK^\mathrm{r}\colon \Omega_0\times
[0,T]\mapsto \mathbb{GL}^3$, where $\mathbb{GL}^3$ is the space of
$3\times 3$ matrices. The reference configuration is $\Omega_0
\subset \mathbb{R}^3$. Our first assumption is that
$\bK^\mathrm{r}$ is given. Future communications will address the
derivation of such an evolution law from our experiments (see
\cite{Calveetal:03} for preliminary results). Since
$\bK^\mathrm{r}$ is generally incompatible (see Section
\ref{sect1}), a further tangent map of material motion,
$\bK^\mathrm{c}\colon \Omega_0\times [0,T]\mapsto \mathbb{R}^3$,
acts to render the tissue of interest, $\mathcal{B}$, compatible
in its remodelled configuration, $\Omega^\ast_t$. The
point-to-point map $\Bkappa\colon\Omega_0\times[0,T] \mapsto
\mathbb{R}^3$ carries material points from $\Omega_0$ to
$\Omega^\ast_t\subset \mathbb{R}^3$, the remodelled configuration.
It is a material motion, and its \emph{compatible} tangent map,
$\bK =
\partial\Bkappa/\partial\bX$ satisfies $\bK =
\bK^\mathrm{c}\bK^\mathrm{r}$. The placement of material points in
$\Omega^\ast_t$ is $\bX^\ast = \Bkappa(\bX,t)$. Further
deformation, brought  about by the displacement, $\bu^\ast$,
carries material points from $\Omega^\ast_t$ to the spatial
configuration $\Omega_t$. The deformation gradient is $\bF^\ast =
{\bf 1} +
\partial\bu^\ast/\partial\bX^\ast$. In this initial treatment we
do not consider any further decompositions of $\bF^\ast$. The
overall motion of a point is $\Bvarphi(\bX,t) = \Bkappa(\bX,t) +
\bu^\ast(\bX^\ast,t)\circ\Bkappa(\bX,t)$, and the corresponding
tangent map is $\bF = {\bf 1} +
\partial\Bvarphi/\partial\bX$. It admits the multiplicative
decomposition $\bF = \bF^\ast\bK^\mathrm{c}\bK^\mathrm{r}$. To
reiterate upon the foregoing distinction between the material
motion and deformation components of the kinematics, we emphasize
that $\bu^\ast$ is a displacement, while $\Bkappa$ is a motion in
material space. The corresponding tangent maps are $\bF^\ast$ (a
classical deformation gradient) and $\bK$ (a material motion
gradient).

\subsection{A variational formulation}\label{sect2.1}

We consider the following energy functional:
\begin{eqnarray}
\Pi[\bu^\ast,\Bkappa] &:=&
\int\limits_{\Omega^\ast_t}\hat{\psi}^\ast(\bF^\ast,\bK^\mathrm{c},\bX^\ast)\mathrm{d}V^\ast\nonumber\\
& &
-\int\limits_{\Omega^\ast_t}\bff^\ast\cdot(\bu^\ast+\Bkappa)\mathrm{d}V^\ast
-\int\limits_{\partial\Omega^\ast_t}\bar{\bt}^\ast\cdot(\bu^\ast+\Bkappa)\mathrm{d}A^\ast,
\label{functnl}
\end{eqnarray}

\noindent where $\psi^\ast =
\hat{\psi}^\ast(\bF^\ast,\bK^\mathrm{c},\bX^\ast)$ is the stored
energy function. Observe that $\psi^\ast$ is assumed to depend
upon the compatibility-restoring material motion,
$\bK^\mathrm{c}$, in addition to the usual dependence on
$\bF^\ast$. Furthermore, material heterogeneity is allowed. The
body force per unit volume in $\Omega^\ast_t$ is $\bff^\ast$, and
the surface traction per unit area on $\partial\Omega^\ast_t$ is
$\bar{\bt}^\ast$. Since the total motion of a material point is
$\Bkappa + \bu^\ast$, the potential energy of the external loads
is as seen in the second and third terms.

Recall that the Euler-Lagrange equations obtained by imposing
equilibrium with respect to $\bu^\ast$ (stationarity of $\Pi$ with
respect to variations in $\bu^\ast$) represent the quasistatic
balance of linear momentum in $\Omega^\ast_t$.
\begin{eqnarray}
& &
\frac{\mathrm{d}}{\mathrm{d}\varepsilon}\Pi[\bu^\ast_\varepsilon,\Bkappa]\Big\vert_{\varepsilon=0}=0\;
\mathrm{where}\; \bu^\ast_\varepsilon = \bu^\ast +
\varepsilon\delta\bu^\ast\label{min1} \\
& &\Longrightarrow \mathrm{Div}^\ast\bP^\ast + \bff^\ast =
\bzero,\;\mathrm{in}\;\Omega^\ast;\; \bP^\ast\bN^\ast =
\bar{\bt}^\ast\;\mathrm{on}\;\partial\Omega^\ast;\;\bP^\ast
:=\frac{\partial\psi^\ast}{\partial\bF^\ast},\label{quasistatic}
\end{eqnarray}

\noindent Observe that the definition of $\bP^\ast$ resembles the
constitutive relation for the first Piola-Kirchhoff stress if
$\Omega^\ast_t$ were the reference configuration.

A final assumption is that the tissue also reaches local
equilibrium with respect to $\Bkappa$ (stationarity of $\Pi$ with
respect to variations in $\Bkappa$). The variational statement is:
\begin{equation}
\frac{\mathrm{d}}{\mathrm{d}\varepsilon}\Pi[\bu^\ast,\Bkappa_\varepsilon]\Big\vert_{\varepsilon=0}=0,
\;\mathrm{where}\, \Bkappa_\varepsilon = \Bkappa +
\varepsilon\delta\Bkappa,\;\mathrm{and}\,\Bvarphi\;\mathrm{is}\;\mathrm{fixed}.
\label{min2}
\end{equation}

\noindent The calculations are lengthy, but entirely standard, and
yield the following Euler-Lagrange equations:
\begin{eqnarray}
-\mathrm{Div}^\ast\left[\psi^\ast{\bf 1} -
\bF^{\ast^\mathrm{T}}\bP^\ast +
\frac{\partial\psi^\ast}{\partial\bK^\mathrm{c}}\bK^{\mathrm{c}^\mathrm{T}}\right]
+ \frac{\partial\psi^\ast}{\partial\bX^\ast} &=& {\bf 0}\;
\mathrm{in} \;\Omega^\ast\label{diffeq}\\
\left[\psi^\ast{\bf 1} - \bF^{\ast^\mathrm{T}}\bP^\ast +
\frac{\partial\psi^\ast}{\partial\bK^\mathrm{c}}\bK^{\mathrm{c}^\mathrm{T}}\right]\bN^\ast
&=& {\bf 0}\;\mathrm{on}\; \partial\Omega^\ast.\label{bouncond}
\end{eqnarray}

Observe that the Eshelby stress $\psi^\ast{\bf 1} -
\bF^{\ast^\mathrm{T}}\bP^\ast$ makes its appearance. Hereafter, it
will be denoted $\mbox{\boldmath$\sE$}$. The term
$\frac{\partial\psi^\ast}{\partial\bK^\mathrm{c}}\bK^{\mathrm{c}^\mathrm{T}}$
is a thermodynamic driving quantity giving the change in stored
energy, $\psi^\ast$, corresponding to a change in configuration,
$\bK^\mathrm{c}$. It is stress-like in its physical dimensions and
tensorial form, and we therefore refer to it as a configurational
stress. Hereafter we will write $\BSigma^\ast =
\frac{\partial\psi^\ast}{\partial\bK^\mathrm{c}}\bK^{\mathrm{c}^\mathrm{T}}$.

\noindent{\bf Remark 1}: A distinct class of variations can be
considered than those in (\ref{min2}). Specifically, consider
\begin{equation}
\frac{\mathrm{d}}{\mathrm{d}\varepsilon}\Pi[\bu^\ast,\Bkappa_\varepsilon]\Big\vert_{\varepsilon=0}=0,
\;\mathrm{where}\, \Bkappa_\varepsilon = \Bkappa +
\varepsilon\delta\Bkappa,\;\mathrm{and}\,\Bvarphi_\varepsilon =
\bX + \Bkappa + \varepsilon\delta\Bkappa + \bu^\ast, \label{min3}
\end{equation}

\noindent which is distinct from (\ref{min2}) in that variations
on the material motion result in variations on the final placement
as well. In this case too, the Euler-Lagrange equations
(\ref{diffeq}) and (\ref{bouncond}) are arrived at. This is an
important property: The system of equations governing the
evolution of the microstructural configuration must be independent
of the particular class of variations considered.

\subsection{Restrictions from the dissipation
inequality}\label{sect2.2}

The dissipation inequality written per unit volume in the
reference configuration takes the familiar form,
\begin{equation}
\Btau\colon(\dot{\bF}\bF^{-1}) - \frac{\partial}{\partial
t}\left(\mathrm{det}[\bK]\psi^\ast\right) \ge 0, \label{dissip1}
\end{equation}

\noindent where $\Btau$ is the Kirchhoff stress defined in
$\Omega_t$. Observe that $\mathrm{det}[\bK]\psi^\ast$ is the
stored energy per unit volume in $\Omega_0$. Using the
multiplicative decomposition $\bF =
\bF^\ast\bK^\mathrm{c}\bK^\mathrm{r}$, and the defining relation
for the configurational stress $\BSigma^\ast =
\frac{\partial\psi^\ast}{\partial\bK^\mathrm{c}}\bK^{\mathrm{c}^\mathrm{T}}$,
standard manipulations result in the following equivalent form:
\begin{eqnarray}
& &
\left(\Btau\bF^{\ast^{-\mathrm{T}}}-\mathrm{det}[\bK]\frac{\partial\psi^\ast}{\partial
\bF^\ast}\right)\dot{\bF}^\ast
-\mathrm{det}[\bK]\left(\mbox{\boldmath$\sE$} +
\BSigma^\ast\right)\colon\left(\dot{\bK}^\mathrm{c}{\bK^\mathrm{c}}^{-1}\right)\nonumber\\
& &-\mathrm{det}[\bK]\mbox{\boldmath$\sE$}
\colon\left(\bK^\mathrm{c}\dot{\bK}^\mathrm{r}\bK^{-1}\right)
-\mathrm{det}[\bK]\frac{\partial\psi^\ast}{\partial\bX^\ast}\cdot\dot{\Bkappa}
\ge 0.\label{dissip2}
\end{eqnarray}

\noindent Adopting the constitutive relation, $\Btau =
\mathrm{det}[\bK]\frac{\partial\psi^\ast}{\partial\bF^\ast}\bF^{\ast^\mathrm{T}}$
(this is consistent with the observation that $\bP^\ast =
\frac{\partial\psi^\ast}{\partial \bF^\ast}$ is related to the
first Piola-Kirchhoff stress with $\Omega^\ast_t$ as the reference
configuration), results in the reduced dissipation inequality
\begin{eqnarray}
& & -\mathrm{det}\bK\left(\mbox{\boldmath$\sE$} +
\BSigma^\ast\right)\colon\left(\dot{\bK}^\mathrm{c}{\bK^\mathrm{c}}^{-1}\right)\nonumber\\
& &-\mathrm{det}[\bK]\mbox{\boldmath$\sE$}
\colon\left(\bK^\mathrm{c}\dot{\bK}^\mathrm{r}\bK^{-1}\right)
-\mathrm{det}[\bK]\frac{\partial\psi^\ast}{\partial\bX^\ast}\cdot\dot{\Bkappa}
\ge 0,\label{dissip3}
\end{eqnarray}
\noindent which places restrictions on the evolution law for
$\bK^\mathrm{r}$, and on the functional dependencies
$\hat{\psi}^\ast(\bullet,\bK^\mathrm{c},\bullet)$ through
$\BSigma^\ast$, and $\hat{\psi}^\ast(\bullet,\bullet,\bX^\ast)$.

\section{Remodelling of one-dimensional bars}\label{sect3}

In general, the examples of remodelling encountered in soft and
hard biological tissue involve complex microstructural changes.
Evolution laws for $\bK^\mathrm{r}$ and the functional form,
$\hat{\psi}^\ast(\bullet,\bK^\mathrm{c},\bullet)$, to model these
complexities are critical components for the successful
application of the theoretical framework outlined in this paper.
In future communications we will describe our experimental program
to extract such constitutive information. However, the working of
the formulation can be demonstrated by academic, but illuminating,
examples. In the interest of brevity we restrict ourselves to a
single example in this paper.
\begin{figure}[ht]
\centering {\includegraphics[width=9cm]{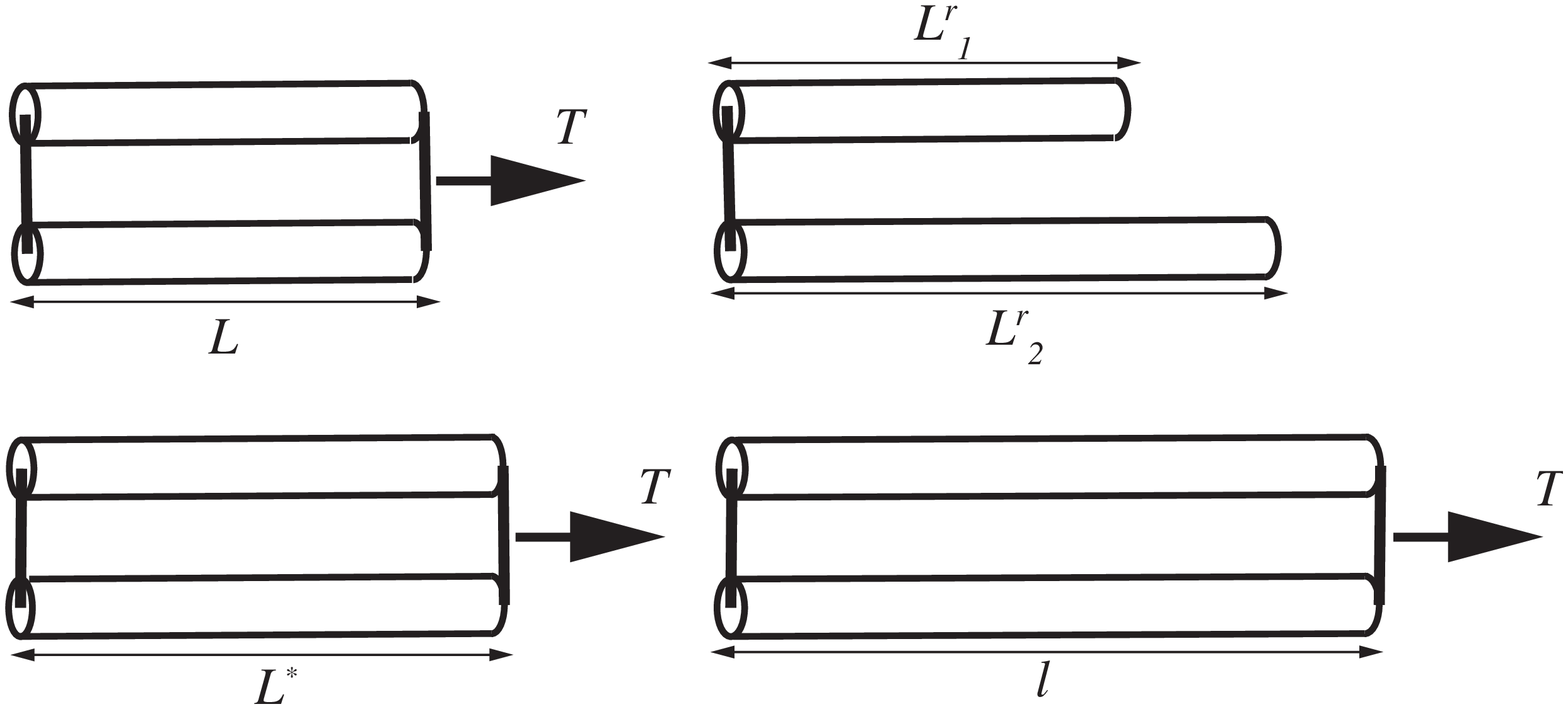}} \caption{\small
One-dimensional remodelling of bars.}\label{fig3}
\end{figure}

Consider two parallel bars, that may represent adjacent strips of
a long bone (Figure \ref{fig3}). We wish to consider a scenario in
which each bar undergoes a preferred material motion to change its
length from $L$ to $L_i^\mathrm{r},\; i=1,2$. In general, this
configuration is incompatible as the bars can attain different
lengths. If they are required to remain of the same length in the
remodelled configuration, further material motion occurs,
resulting in a length $L^\ast$ for each bar. This is the
remodelled configuration, $\Omega^\ast_t$, in which the total
material motion of each bar is $\kappa = L^\ast - L$. If the
remodelling takes place under an external load, $T$, the bars each
stretch to a final length $l$. The deformation is $u^\ast = l -
L^\ast$. We examine the equations that govern the deformation and
material motion by considering the following energy functional:
\begin{equation}
\Pi[u^\ast,\kappa] = \frac{1}{2}k^\ast(\kappa + L -
L^\mathrm{r}_1)^2 + \frac{1}{2}k^\ast(\kappa + L -
L^\mathrm{r}_2)^2 + 2\cdot\frac{1}{2}k{u^\ast}^2 -
T(u^\ast+\kappa),\label{one-dim-energy}
\end{equation}

\noindent where $k^\ast$ and $k$ are spring constants for the
material motion- and stretch-dependent portions of the stored
energy, respectively. These portions are assumed to be separable.
The theory of Section \ref{sect2} results in the following
relations:
\begin{eqnarray}
& &\frac{\partial\Pi}{\partial u^\ast} = 0\quad \Rightarrow\;
2ku^\ast = T,\label{one-dim-linmom}\\
& &\frac{\partial\Pi}{\partial\kappa} = 0\quad \Rightarrow\;
\Bkappa = \frac{k}{k^\ast}u^\ast - \left(L - \frac{L^\mathrm{r}_1
+ L^\mathrm{r}_2}{2}\right).\label{one-dim-remod}
\end{eqnarray}

\noindent In (\ref{one-dim-linmom}) the standard relation is seen
for the stretch of a linear spring with effective stiffness $2k$.
The more interesting result is (\ref{one-dim-remod}). Observe that
when $L = \frac{1}{2}(L_1^\mathrm{r} + L_2^\mathrm{r})$, material
motion can occur, driven by stress, since $u^\ast = T/2k$ from
(\ref{one-dim-linmom}). In this case remodelling can be
incompatible if $L_1^\mathrm{r} \neq L_2^\mathrm{r}$. However,
remodelling does not drive material motion, $\Bkappa$ in this
case. Instead there is stress-driven remodelling as described in
Section \ref{sect1}. On the other hand, in the absence of an
external load, material motion is obtained when $L \ne
\frac{1}{2}(L_1^\mathrm{r} + L_2^\mathrm{r})$. In this case the
compatibility-restoring remodelling, motivated in Section
\ref{sect1} and described by the tangent map, $\bK^\mathrm{c}$,
also leads to overall material motion, $\Bkappa$.

\section{Discussion and conclusion}\label{sect4}

This paper has presented a theoretical framework for remodelling
in biological tissue, where this phenomenon is understood as an
evolution of the microstructural configuration of the material.
The assumption that the material attains local equilibrium with
respect to the evolution of its microstructure results in
Euler-Lagrange relations in the form of a governing partial
differential equation and boundary conditions. The final form of
the equations and the results themselves depend critically upon
the specified constitutive relations for the preferred remodelled
state of the material, and the dependence of the stored energy
upon the compatibility-restoring component of remodelling. These
are open problems, and will be addressed in future papers by our
group. The following points are noteworthy at this stage of the
development of the theory:
\begin{itemize}
\item This work does not deal with the approach to local
equilibrium, which may take time on the order of days in
biological tissue. Furthermore, the equilibrium state, being
defined by the external loads, evolves upon perturbation of these
conditions: Additional remodelling occurs in biological tissue
when the load is altered.
\item The energy functionals in (\ref{functnl}) and
(\ref{one-dim-energy}) generalize to the Gibbs free energy of the
body under constant loads and isothermal conditions. Further
contributions that drive the process, such as chemistry or
electrical stimuli, can be encompassed by the Gibbs energy. In
such a setting the process we have described here would be termed
a ``self-assembly'' in the realms of materials science or physics.
\end{itemize}

\bibliographystyle{elsart-harv}
\bibliography{mybib}

\begin{thebibliography}{5}
\expandafter\ifx\csname natexlab\endcsname\relax\def\natexlab#1{#1}\fi
\expandafter\ifx\csname url\endcsname\relax
  \def\url#1{\texttt{#1}}\fi
\expandafter\ifx\csname urlprefix\endcsname\relax\def\urlprefix{URL }\fi

\bibitem[{Ambrosi and Mollica(2002)}]{AmbrosiMollica:02}
Ambrosi, D., Mollica, F., 2002. {On the mechanics of a growing tumor}. {Int. J.
  Engr. Sci.} 40, 1297--1316.

\bibitem[{Calve et~al.(2003)Calve, Dennis, Kosnik, Baar, Grosh, and
  Arruda}]{Calveetal:03}
Calve, S., Dennis, R., Kosnik, P., Baar, K., Grosh, K., Arruda, E., 2003.
  {Engineering of functional tendon}, submitted to {\sl Tissue Engineering}.

\bibitem[{Humphrey and Rajagopal(2002)}]{HumphreyRajagopal:02}
Humphrey, J.~D., Rajagopal, 2002. {A constrained mixture model for growth and
  remodeling of soft tissues}. {M}ath. {M}eth. {M}od. {A}pp. {S}ci. 12~(3),
  407--430.

\bibitem[{Taber(1995)}]{Taber:95}
Taber, L.~A., 1995. {Biomechanics of growth, remodelling and morphogenesis}.
  Applied {M}echanics {R}eviews 48, 487--545.

\bibitem[{Taber and Humphrey(2001)}]{TaberHumphrey:01}
Taber, L.~A., Humphrey, J.~D., 2001. {Stress-modulated growth, residual stress
  and vascular heterogeneity}. {J.} {B}io. {M}ech. {E}ngrg. 123, 528--535.

\end{thebibliography}

\end{document}